\documentclass[journal=apchd5,manuscript=letter]{achemso}
\usepackage[version=3]{mhchem} 

\let\vec\mathbf
\newcommand{\RNum}[1]{\uppercase\expandafter{\romannumeral #1\relax}}
\author{Minkyung Kim}
\altaffiliation{These authors contributed equally to this work.}
\author{Dasol Lee}
\affiliation[ME]
{Department of Mechanical Engineering, Pohang University of Science and Technology (POSTECH), Pohang 37673, Republic of Korea}
\altaffiliation{These authors contributed equally to this work.}
\author{Tae Hak Kim}
\affiliation[Ajou1]
{Department of Energy Systems Research, Ajou University, Suwon 16499, Republic of Korea}
\author{Younghwan Yang}
\affiliation[ME]
{Department of Mechanical Engineering, Pohang University of Science and Technology (POSTECH), Pohang 37673, Republic of Korea}
\author{Hui Joon Park}
\affiliation[Ajou1]
{Department of Energy Systems Research, Ajou University, Suwon 16499, Republic of Korea}
\alsoaffiliation[Ajou2]
{Department of Electrical and Computer Engineering, Ajou University, Suwon 16499, Republic of Korea}
\author{Junsuk Rho}
\affiliation[ME]
{Department of Mechanical Engineering, Pohang University of Science and Technology (POSTECH), Pohang 37673, Republic of Korea}
\alsoaffiliation[CE]
{Department of Chemical Engineering, Pohang University of Science and Technology (POSTECH), Pohang 37673, Republic of Korea}
\email{jsrho@postech.ac.kr}

\title[An \textsf{achemso} demo]
  {Observation of enhanced optical spin Hall effect in a vertical hyperbolic metamaterial}

\abbreviations{OSHE,hHMM,vHMM,EMT,RCP,LCP}

\begin{document}


\begin{abstract}
Hyperbolic metamaterials, horizontally stacked metal and dielectric multilayer, have recently been studied as a platform to observe optical spin Hall effect. However, the large optical spin Hall effect in the horizontal hyperbolic metamaterials accompanies extremely low transmission, which obstructs its practical applications. Reducing the sample thickness to augment the transmission causes diminishment of the shift. In this letter, we demonstrate that a vertical hyperbolic metamaterial can enhance the shift by several orders of magnitude in comparison to the shift of its horizontal counterpart. Under the same conditions of material combinations and total thickness, the shift enhancement, which is incident angle-dependent, can be higher than 800-fold when the incident angle is 5$^\circ$, and 5000-fold when the incident angle is 1$^\circ$. As a proof of concept, we fabricate a large-scale gold nano-grating by nanoimprint lithography and measure the helicity-dependent shift by Stokes polarimetry setup, which agrees well with the simulated result. The gigantic optical spin Hall effect in a vertical hyperbolic metamaterial will enable helicity-dependent control of optical devices including filters, sensors, switches and beam splitters.
\end{abstract}

\section{Introduction}
Hyperbolic metamaterials are artificially structured materials that are designed to exhibit a hyperbolic shape of equi-frequency contour. Highly anisotropic electric responses of the hyperbolic metamaterials, represented as opposite signs of permittivities along direction, have shown potential in a variety of applications such as super-resolution imaging \cite{Jacob:06, rho2010spherical, doi:10.1021/acsphotonics.7b01182, byun2017demonstration}, enhancement of spontaneous emission \cite{Jacob2010} and anomalous scaling laws \cite{yang2012experimental}. Among the many applications of hyperbolic metamaterials, optical spin Hall effect (OSHE) has received growing attention recently \cite{7970110, tang2016enhanced, Takayama:18}. The extreme anisotropic properties of the hyperbolic metamaterials facilitate spin-dependent splitting of linearly polarized incidence and thereby amplify OSHE even with a sample that is thinner than a wavelength. OSHE in hyperbolic metamaterials has been recently reported theoretically \cite{7970110, tang2016enhanced} and experimentally \cite{Takayama:18} by horizontally stacking metal and dielectric multilayer alternately, parallel to the normal vector of the substrate; this structure will be referred as a horizontal hyperbolic metamaterial (hHMM). An alternative arrangement is a vertical hyperbolic metamaterial (vHMM), in which interfaces between dielectric and metal are perpendicular to the normal vector of the substrate. 

Here we present several orders of magnitude enhancement of OSHE compared to that in hHMMs using a vHMM. Despite the advantages of vHMMs over hHMMs in many applications such as super-resolution imaging \cite{sun2015experimental} and broadband negative refraction due to the non-resonant characteristics, OSHE using vHMM has been not studied yet. We demonstrate numerically that the different optic axis configuration in vHMMs increases the difference between transmission coefficients of s- and p-polarization, thereby giving rise to significant enhancement of the OSHE. Then, the enhanced OSHE is experimentally verified by a Stokes polarimetry method.

\section{Results and discussion}
OSHE, as a photonic counterpart of the spin Hall effect, refers to the spin-dependent transverse shift of a linearly polarized incident beam \cite{PhysRevLett.93.083901, PhysRevLett.96.073903, Yin1405}. The underlying principle of OSHE is angular momentum conservation, which leads to spin-orbit coupling. The interaction between the circular polarization (spin angular momentum) and the trajectory (orbital angular momentum) results in the deflection of a circularly polarized transmitted beam, in which the transverse shift depends on the handedness of the polarization.

\begin{figure}[htp] \centering
\includegraphics [width=0.7\textwidth]{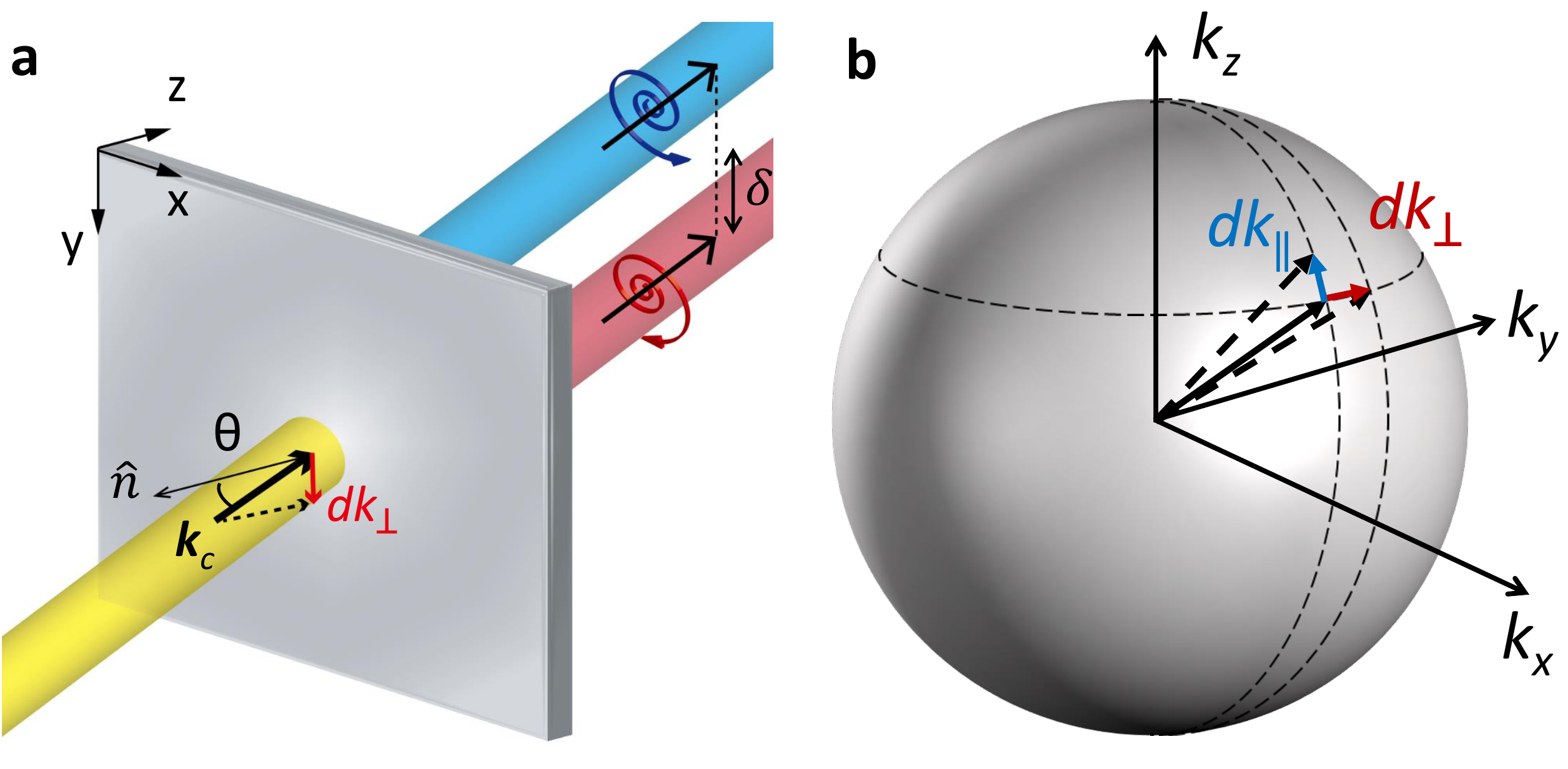}
\caption{Schematics of optical spin Hall effect in (a) real space and (b) momentum space. (a) Incident Gaussian beam with the finite thickness containing non-central wave vectors. Sold arrow: central wave vector $\vec{k}_c$; dashed arrow: out-of-plane wave vector deflected by $d\vec{k}$. The transverse shift is denoted as $\delta$. (b) A sphere in momentum space with in-plane ($dk_\parallel$) and out-of-plane ($dk_\perp$) deflection.
}
\label{schematics}
\end{figure}

When an incident Gaussian beam propagates in the $x$-$z$ plane (Fig. \ref{schematics}(a)), the transmitted beam should ideally also be in the $x$-$z$ plane. However, a Gaussian beam with finite thickness includes non-central wave vectors $\vec{k}_c + d\vec{k}$ where $\vec{k}_c$ is a central wave vector. The deflection $d\vec{k}$ can be either in-plane ($dk_\parallel$) or out-of-plane ($dk_\perp$), which results in discrepancy of incident angle and incident plane respectively (Fig. \ref{schematics}(b)). Considering that the p- and s-polarization are defined with respect to the wave vector, the presence of $d\vec{k}$ also alters the basis of polarization and induces spin-dependent Berry phase \cite{Bliokh_2013}. Therefore, a linearly polarized incidence, which can be regarded as a superposition of left/right-circularly polarized light (LCP/RCP), is split transversely into two circularly polarized beams with the oppositely shifted beam centroids.


The shift is usually limited to a fraction of wavelength, as a result of the weak spin-orbit coupling, and is difficult to observe experimentally. Therefore, observation of the weak OSHE requires amplification of the signal by using quantum weak measurements involving preselection and postselection techniques \cite{Hosten787, PhysRevA.85.043809}. Alternatively, the shift can be enhanced by exploiting recent advances in metamaterials. Enhancement of OSHE based on metamaterials has been demonstrated during the past several years using metasurfaces \cite{Yin1405}, epsilon-near-zero metamaterial \cite{Zhu:15} and hyperbolic metamaterials \cite{7970110, tang2016enhanced, Takayama:18}. Although hHMMs have shown a huge shift of up to a few tens of micrometers \cite{7970110, tang2016enhanced} and even beyond \cite{Takayama:18} in the visible range, they suffer from extremely low transmittance below 0.01, which hinders practical applications. The low efficiency can be overcome by decreasing the thickness, but this change also reduces the shift.

\begin{figure}[htp] \centering
\includegraphics [width=0.9\textwidth]{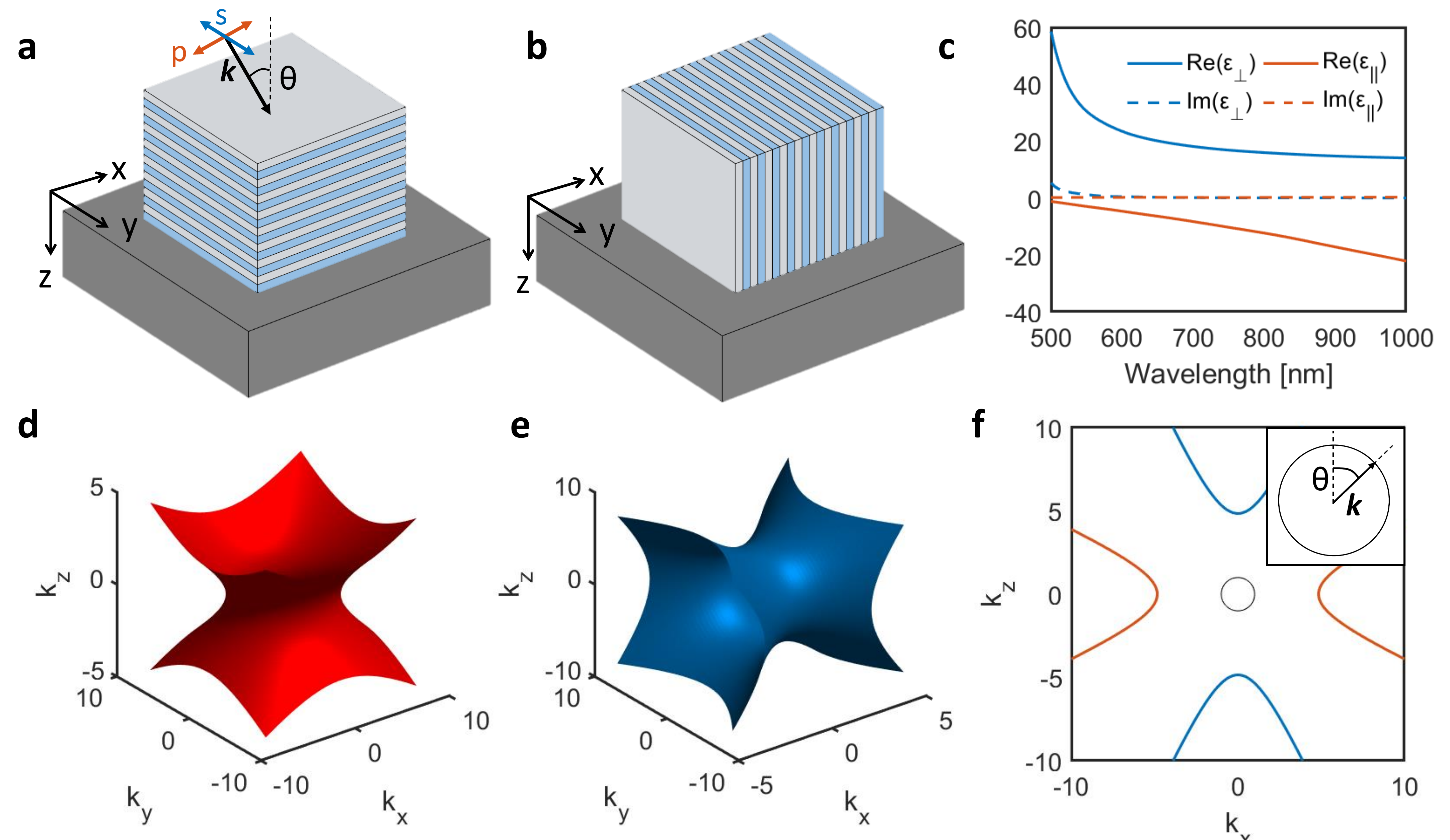}
\caption{Schematics of (a) hHMM and (b) vHMM. (c) Effective permittivities of Ag and TiO\textsubscript{2} multilayer with equal volume fraction. Equi-frequency contour of a (d) hHMM and (e) vHMM consisting of Ag and TiO\textsubscript{2}. (f) Equi-frequency curve of the hHMM (red) and vHMM (blue) in $k_x$-$k_z$ plane. Black curve represents equi-frequency curve of air. Inset shows magnified image of light cone.
}
\label{dispersion}
\end{figure}

We propose a deep-subwavelength thin vHMM, which exhibits a gigantic OSHE with high efficiency, to solve this issue. The optic axis of a vHMM is perpendicular to the normal vector of the substrate, whereas the optic axis of a hHMM is parallel to it. This different configuration enables enhancement of the shift with high efficiency. Recently, OSHE of a beam reflected by black phosphorus, an anisotropic two-dimensional atomic crystal, has been numerically demonstrated \cite{Zhang:18}. It demonstrates the possibility of large OSHE in an atomically thin sample, but the shift barely exceeds one wavelength. 
To enhance the OSHE, we examine the analytic formula of the shift. The transverse shift of the horizontally polarized beam injected from medium 1 with refractive index $n_1$ is provided as \cite{7970110, tang2016enhanced}
\begin{equation}
    \label{shift}
    \delta^{\pm}_{H} =\pm\frac{ k_{1}w_{0}^2\cot{\theta_{i}}
    \big(
    t_{p}^2\frac{\cos{\theta_{t}}}{\cos{\theta_{i}}}-t_{p}t_{s}
    \big)}
    {k_{1}^2w_{0}^2t_{p}^2
    +\cot{\theta_{i}}^2
    \big(
    t_{p}\frac{\cos{\theta_{t}}}{\cos{\theta_{i}}}-t_{s}
    \big)^2
    +(\frac{\partial t_{p}}{\partial\theta_{i}})^2}
\end{equation}
where $k_{1} = n_{1}k_{0}$ where $k_{0}$ is the wave vector in free space; +(-) corresponds to LCP (RCP); $w_{0}$ is a beam waist; $\theta_{i}$ and $\theta_{t}$ are incident and transmitted angle respectively. $t_p$ and $t_s$ are transmission coefficients of p- and s-polarization respectively. When the beam waist is sufficiently large ($k_1^2w_0^2\gg\cot^2{\theta_i}$), the Eq. \ref{shift} can be simplified to \cite{zhou2014observation}
\begin{equation}
    \label{simple_shift}
    \delta^{\pm}_{H} = \pm\frac{\cot{\theta_{i}}}{k_1}
    \text{Re}(1-\frac{t_s}{t_p})
\end{equation}
Since the shift is proportional to the real part of $1-{t_s}/{t_p}$ and $\cot{\theta_i}$, a natural way to enhance the shift is to either increase the difference between $t_s$ and $t_p$, or to decrease $\theta_{i}$. To determine why vHMM amplifies OSHE more than hHMM does, we examine the effective parameters of the two cases. According to effective medium theory (EMT), a structure with a deep-subwavelength scale can be approximated by a homogeneous medium with effective parameters. The permittivity of a multilayer-based hyperbolic metamaterial is \cite{AGRANOVICH198585}
\begin{align}
    \varepsilon_\parallel = f \varepsilon_m + (1-f) \varepsilon_d
    \notag \\
    \varepsilon_\perp = \frac{\varepsilon_m \varepsilon_d}{(1-f) \varepsilon_m + f \varepsilon_d}
\end{align}
where $\varepsilon$ is a relative permittivity, subscript $m$ and $d$ represent metal and dielectric respectively, subscript $\perp$ and $\parallel$ correspond to direction perpendicular and parallel to the interfaces, and $f$ is a metal filling ratio. Permittivity tensor of hHMM is $\varepsilon_\text{hHMM}=\text{diag}(\varepsilon_\parallel, \varepsilon_\parallel, \varepsilon_\perp)$ for hHMM (Fig. \ref{dispersion}(a)) and $\varepsilon_\text{vHMM}=\text{diag}(\varepsilon_\perp, \varepsilon_\parallel, \varepsilon_\parallel)$ for vHMM (Fig. \ref{dispersion}(b)). s-Polarized incidence, whose electric field has a $y$-component only, experiences the y-component of permittivity. Since hHMM and vHMM have the same permittivity along $y$-axis, they have the same $t_s$. In contrast, p-polarized light has both $x$- and $z$-components of the electric field. When incident angle is large ($\theta_i \sim \pi/2$), the p-polarized light experiences the $z$-component of permittivity dominantly. Then, the difference between $t_s$ and $t_p$ of an hHMM is large but the shift remains small due to the $\cot{\theta_i}$ term in Eq. \ref{simple_shift}. However, when incident angle is small ($\theta_i \ll \pi/2$), the $x$-component of permittivity is dominant. Therefore, a vHMM exhibits significant discrepancy between $t_p$ and $t_s$ at small incident angle while they are rather similar in an hHMM. The polarization-dependent transmission coefficients of vHMMs at small $\theta_i$ yield a gigantic OSHE. This EMT-based analysis can be applied to general anisotropic materials, not limited to hyperbolic metamaterials. Therefore, a vertically stacked dielectric multilayer, which possesses elliptical equi-frequency contour, can also induce higher OSHE compared to their horizontal counterpart. However, due to weak anisotropy in the dielectric multilayer, its OSEH is not as strong as in hyperbolic metamaterials (See supplementary).

The enhanced OSHE in vHMM can be also understood by examining an equi-frequency contour. To compare the difference between hHMM and vHMM clearly, we use a multilayer of silver (Ag) and titanium dioxide (TiO\textsubscript{2}) with $f=0.5$ (Fig. \ref{dispersion}(a) and \ref{dispersion}(b)), which has positive $\varepsilon_\perp$ and negative $\varepsilon_\parallel$ (Fig. \ref{dispersion} (c)). Such a medium with two negative and one positive diagonal element of permittivity tensor exhibits type-\RNum{2} hyperbolic dispersion. Equi-frequency contours of hHMM and vHMM at 600 nm are shown in Fig. \ref{dispersion}(d)-\ref{dispersion}(f). The hyperbolic equi-frequency contours correspond to p-polarized modes. On the contrary, the s-polarized modes have purely imaginary equi-frequency contours originating from the negative permittivity elements. Therefore, incident light with s-polarization cannot be effectively coupled to the hyperbolic metamaterials. It is also true for p-polarized light incident to hHMM. Considering that the tangential wave vector $k_x$ should be continuous at the interface, the light cone cannot excite any mode of the hHMM, because p-polarized mode of the hHMM has high $k_x$-components only (Fig. \ref{dispersion}(f)). In contrast, because the optic axis of a vHMM is along the $x$-axis, the vHMM has p-polarized $k_x$-components which overlap those of the light cone. Therefore, p-polarized incidence can couple to vHMM, which gives rise to higher transmission of vHMMs in comparison to hHMMs.


\begin{figure}[htp] \centering
\includegraphics [width=0.99\textwidth]{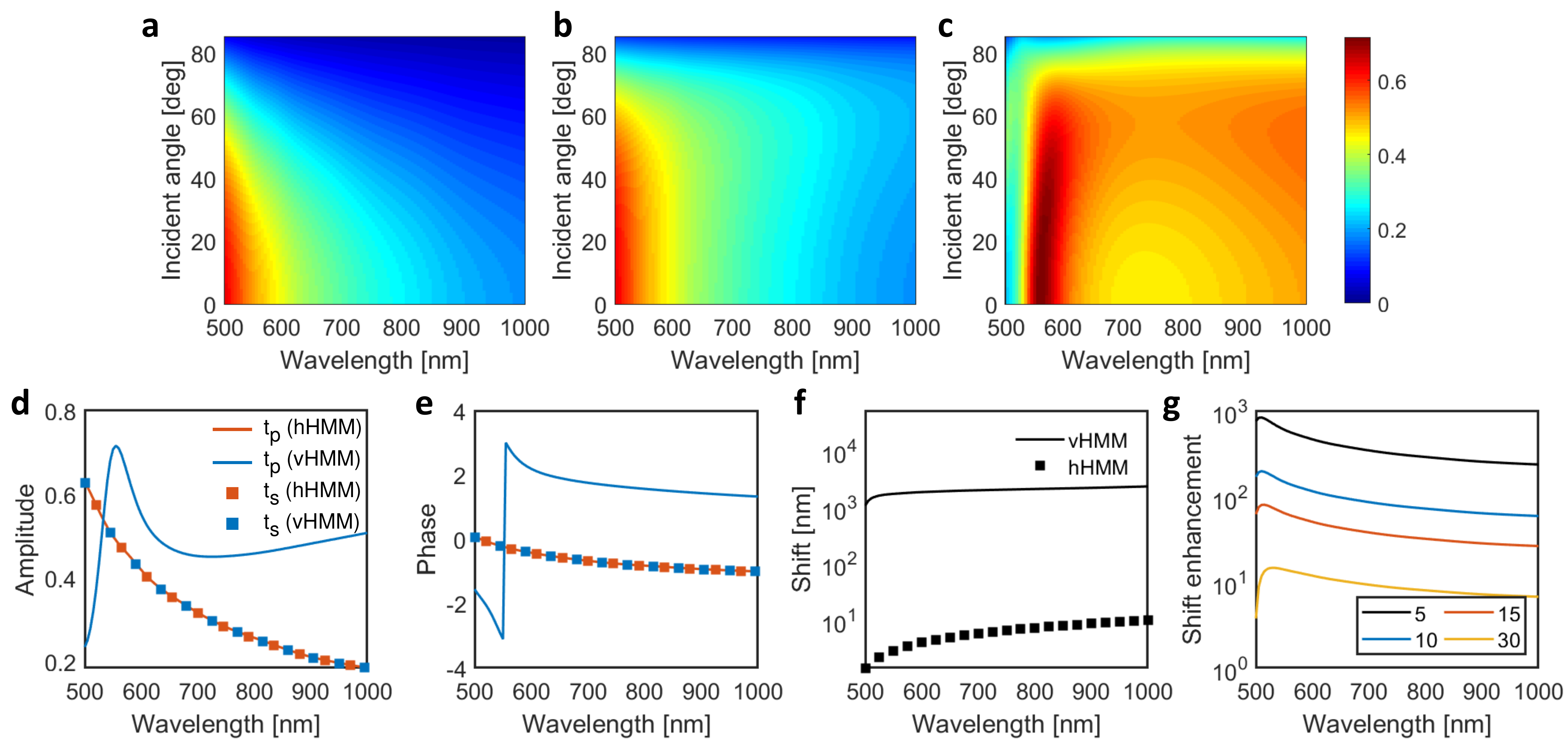}
\caption{Amplitude of transmission coefficients of (a) s- and (b) p-polarization of hHMM. (c) Amplitude of transmission coefficient of p-polarization of vHMM. (d) Amplitude and (e) phase of transmission coefficients and (f) shift when incident angle is 5$^{\circ}$. (g) Shift enhancement ($\delta_\text{vHMM}/\delta_\text{hHMM}$) calculated by Eq. \ref{simple_shift} at four different incident angle.
}
\label{coefficient}
\end{figure}

In order to quantitatively compare the transmission coefficients of hHMM and vHMM, $t_s$ and $t_p$ of two media are calculated using EMT (See supplementary). Transmission coefficients of an hHMM consisting of Ag and TiO\textsubscript{2} layers with total thickness $d=50$ nm placed on a glass substrate are shown in Fig. \ref{coefficient}(a) and \ref{coefficient}(b). Because $t_s$ of hHMM and vHMM is the same, only $t_p$ of a vHMM composed of the same materials and total thickness is shown in Fig. \ref{coefficient}(c). Difference of $t_s$ and $t_p$ is noticeable in the vHMM while they are unremarked in the hHMM especially at small $\theta_i$. To emphasize the difference, transmission coefficients and shift when $\theta_i = 5^{\circ}$ are plotted in Fig. \ref{coefficient}(d)-\ref{coefficient}(f). Shifts $\delta$ calculated based on Eq. \ref{simple_shift} show that despite the deep-subwavelength total thickness, the vHMM supports micrometer-scale transverse shift, which corresponds to several wavelengths, whereas the hHMM achieves a shift of only a few nanometers. Shift enhancement, which is defined as the shift in vHMM divided by the shift in hHMM ($\delta_\text{vHMM}/\delta_\text{hHMM}$), proves significant enhancement in a broad range. The shift enhancement, which depends on $\theta_i$, shows extreme enhancement at small $\theta_i$ and therefore can be further increased by decreasing it (Fig. \ref{coefficient}(g)). The shift is 800-fold enhanced when $\theta_i = 5^{\circ}$ (Fig. \ref{coefficient}(f)), and 5000-fold enhanced when $\theta_i = 1^{\circ}$ (See supplementary). We would like to emphasize that although we use Ag and TiO\textsubscript{2} for quantitative demonstration, the enhancement of OSHE in vHMM is true in general, and is not limited to the specific combination of materials.

\begin{figure}[htp] \centering
\includegraphics [width=0.99\textwidth]{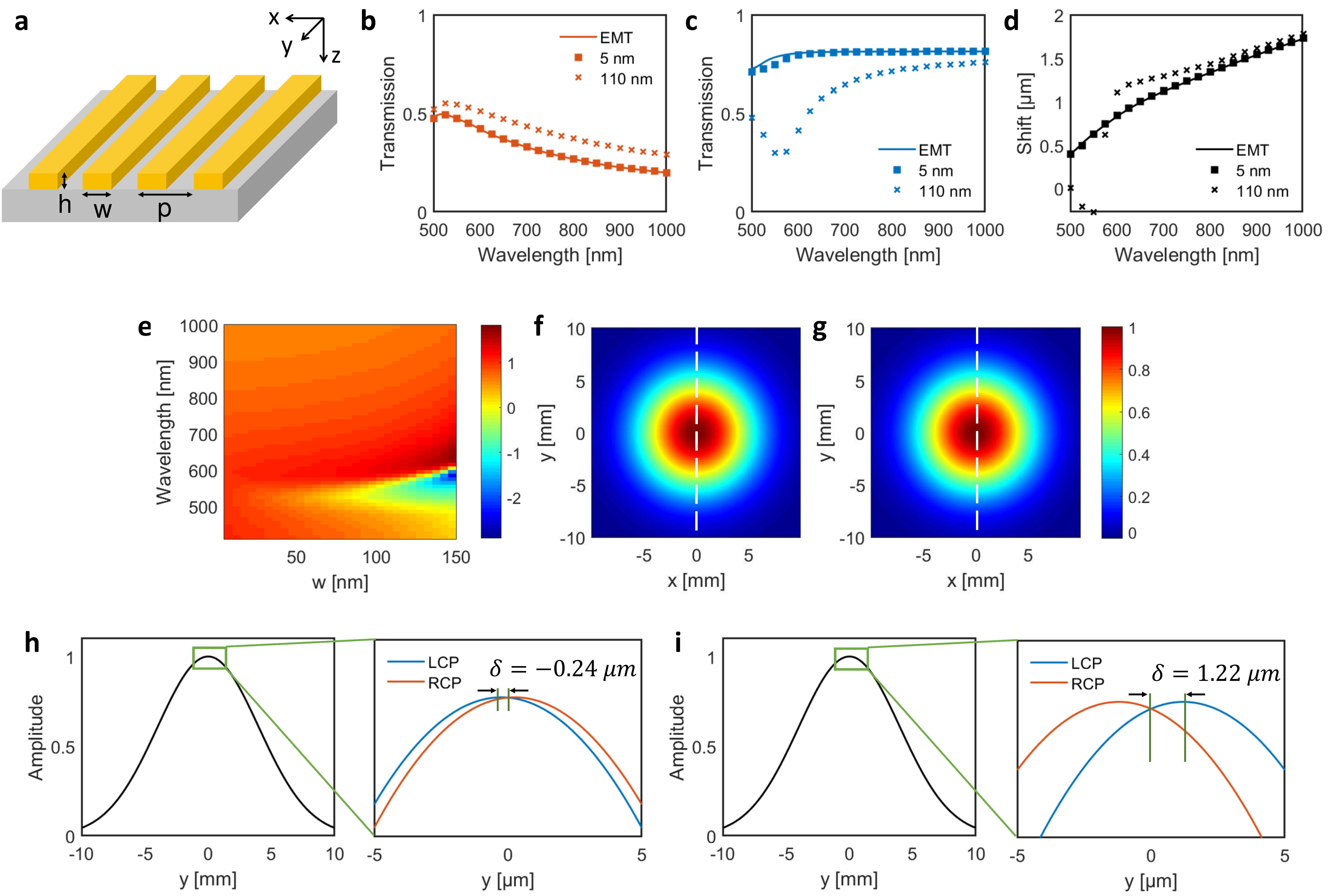}
\caption{(a) A schematic of realistic vHMM. Amplitude of transmission coefficient of (b) s- and (c) p-polarization. (d) Shift calculated by EMT and realistic structure. (e) Shift of the realistic structure divided by the shift calculated by EMT. Field profile of (f) left- and (g) right-circularly polarized light at 532 nm. Field amplitude along the white dashed line at (h) 532 nm and (i) 638 nm. Beam waist $w_0$ is 4 mm, and $\theta_i = 5^{\circ}$.
}
\label{field}
\end{figure}

While hHMMs can be fabricated readily by repetition of thin film evaporation, fabrication of vHMM is challenging because of high aspect ratio and difficulty in filling the vacancies. Therefore, we propose a gold nano-grating on a glass substrate as a realistic structure (Fig. \ref{field}(a)). Here, air ($n=1$) corresponds to the dielectric part, and the effective medium consisting of gold and air with $f=0.5$ possesses type-\RNum{2} hyperbolic dispersion (See supplementary). The geometrical parameters are taken as: width $w = 110$ nm, period $p = 2w$ and height $h = 50$ nm. Despite the thickness of a few tens of nanometers, such grating structures have shown hyperbolic dispersion and related phenomena such as negative refraction and diffraction-unlimited propagation of surface waves \cite{PhysRevLett.97.073901, doi:10.1063/1.4821444, high2015visible}.

Transmission coefficients of the gold nano-grating are calculated using COMSOL Multiphysics (cross markers in Fig. \ref{field}(b) and \ref{field}(c)). Transmission coefficients when $w = 5$ nm and those calculated by EMT are plotted as references. Then we use Eq. \ref{simple_shift} to calculate the transverse shift. Since the realistic structure is not sufficiently deep-subwavelength, the spectra and shift disagree slightly with the result obtained using EMT. Shift of the realistic structure divided by the shift calculated by EMT is shown in Fig. \ref{field}(e). The geometrical discrepancy between the realistic structure and the effective medium model results in disagreement at short wavelength. As the width of the grating becomes comparable to the wavelength, EMT breaks down, and the reflection spectrum of the grating exhibits a broad peak near 600 nm for p-polarization. The high reflection results in a dip in $t_p$, which is not predicted by EMT. Nevertheless, shift of the realistic structure is still several orders of magnitude higher than that of hHMMs.


Since the transmission coefficients are given, we can calculate field amplitude distribution of a circularly polarized transmitted Gaussian beam by applying the boundary conditions \cite{tang2016enhanced}. The field distributions of the transmitted beam at 532 nm are shown in Fig. \ref{field}(f) and \ref{field}(g). Field amplitude along the white dashed line shows transverse shift of two circularly polarized light. At 532 nm where the shift is negative, LCP (RCP) is shifted to the -$y$ (+$y$) direction with $\delta=-0.24$ $\mu$m (Fig. \ref{field}(h)). On the other hand, at 638 nm where the shift is positive, the shift is reversed and LCP is shifted by $\delta=1.22$ $\mu$m (Fig. \ref{field}(i)).

\begin{figure}[htp] \centering
\includegraphics [width=0.95\textwidth]{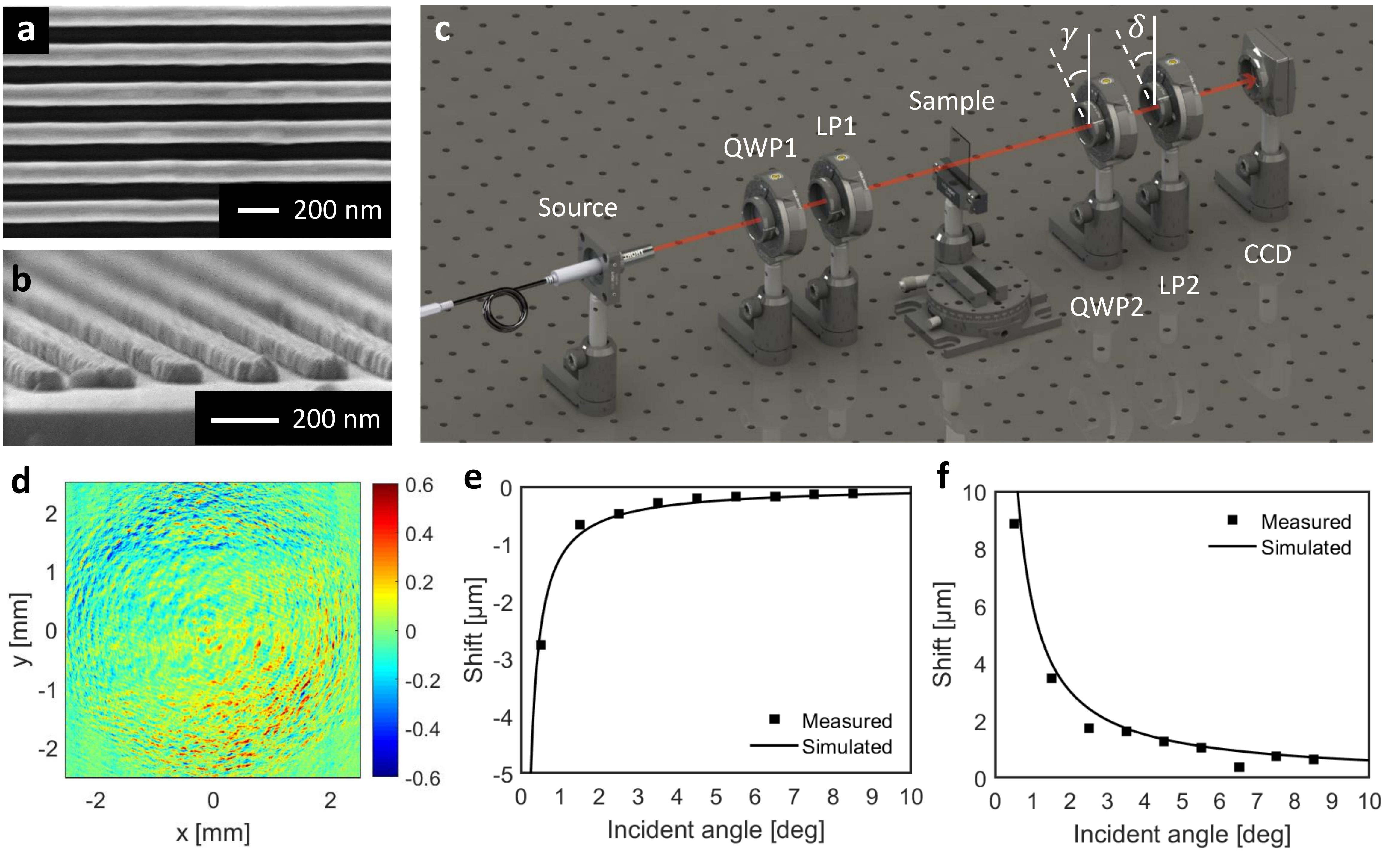}
\caption{(a) Top view and (b) perspective view of the fabricated sample obtained by scanning electron microscopy. (c) Schematics of Stokes polarimetry setup. (d) Experimentally measured $s_3$ distribution at 532 nm when the incident angle is 5.5$^{\circ}$. Simulated and measured shift at (e) 532 nm and (f) 638 nm. Error bars are not plotted here as they are smaller than the point marker.
}
\label{experiment}
\end{figure}

A centimeter-scale gold nano-grating is patterned on a glass substrate by using nanoimprint lithography. Scanning electron microscopy images of the fabricated sample are shown in Fig. \ref{experiment}(a) and \ref{experiment}(b). In order to measure the spin-dependent transverse shift, we measure the Stokes parameters, which characterize the polarization states, of the transmitted beam by using a polarimetry setup (Fig. \ref{experiment}(c)) \cite{Bliokh:16,Takayama:18}. Diode lasers with wavelength 532 nm and 638 nm are used as sources. The incident Gaussian beam is linearly polarized by 45$^{\circ}$ after passing the first linear polarizer (LP1). Then the beam passes the sample, a quarter-wave plate and an additional linear polarizer in sequence and then is captured by a position-sensitive CCD camera. The distributions of Stokes parameters of the transmitted beam are \cite{Bliokh:16}
\begin{align}
    S_0(\vec{R}) &= I(0^{\circ},0^{\circ}) + I(0^{\circ},90^{\circ}) \notag \\
    S_1(\vec{R}) &= I(0^{\circ},0^{\circ}) - I(0^{\circ},90^{\circ}) \notag \\
    S_2(\vec{R}) &= I(0^{\circ},45^{\circ}) - I(0^{\circ},135^{\circ}) \notag \\
    S_3(\vec{R}) &= I(90^{\circ},45^{\circ}) - I(90^{\circ},135^{\circ})
\end{align}
where $I(\delta,\gamma)$ is the local intensity when retardation angle of the second quarter-wave plate (QWP2) is $\delta$ and rotation angle of the second linear polarizer (LP2) is $\gamma$. The normalized third stokes parameter $s_3(\vec{R}) = S_3(\vec{R})/S_0(\vec{R})$ represents the local ellipticity of the transmitted Gaussian beam (Fig. \ref{experiment}(d)). Different handedness of the ellipticity manifests the helicity-dependent splitting along the $y$-axis. Red and blue color in Fig. \ref{experiment}(d) correspond to LCP and RCP respectively. As predicted by Fig. \ref{field}, LCP (RCP) is shifted to -$y$ (+$y$) direction at 532 nm. Furthermore, we calculated the shift from the measured Stokes parameter using the following formula \cite{Bliokh:16}.

\begin{equation}
    \delta = \frac{\cot{\theta_i}}{k}(-\sigma(1-\cos{\Phi_0}) + \chi \sin{\Phi_0})
    \label{measured_shift}
\end{equation}

Here, $\Phi_0 = \tan^{-1}(\tilde{S_3}/\tilde{S_2})$ where $\tilde{S_j}$ is the integrated Stokes parameter defined as $\tilde{S_j} = \int{S_j(\vec{R}) d^2\vec{R}}$, $\sigma=2 \text{Im}(\alpha^* \beta)$ and $\chi = 2 \text{Re}(\alpha^* \beta)$ where $(\alpha \hspace{3mm} \beta)^T$ is the Jones vector of the incident beam ($\alpha = \beta = 1/\sqrt{2}$). The measured shift at 532 nm (Fig. \ref{experiment}(e)) and at 638 nm (Fig. \ref{experiment}(f)) obtained by Eq. \ref{measured_shift} agree well with the simulation. In ten consecutive experiments performed with the same incident angle, $\delta$ and $\gamma$, the standard deviation of the shift is less than one percent of its value. As predicted by Eq. \ref{simple_shift}, the shift diverges as the incident angle goes to zero. The maximum shift, which is approximately 9 $\mu$m in our measurement can be further enhanced significantly by reducing $\theta_i$. The measured transmittance of the sample, which is calculated by the ratio of $\tilde{S_0}$ of the sample to $\tilde{S_0}$ of free space, is $0.44 \pm 0.01$ at 532 nm and 0.25 at 632 nm. The high efficiency resulting from a deep-subwavelength thin sample will be beneficial in developing spin-dependent photonic devices.

In conclusion, we demonstrate an enhanced optical spin Hall effect in a deep-subwavelength thin vertical hyperbolic metamaterial. High discrepancy between the effective permittivities experienced by s- and p-polarization provides the several orders of magnitude enhanced shift compared to that in a horizontal hyperbolic metamaterial. a micrometer-scale transverse shift by using a gold nano-grating is experimentally demonstrated by Stokes polarimetry. The huge optical spin Hall effect in a vertical hyperbolic metamaterial provides a way to realize compact photonic devices with spin degree-of-freedom such as filters, sensors, switches and beam splitters.

\section{METHODS}

\subsection{Fabrication}
Gold nano-grating patterns (2 $\times$ 2 cm area) having 220 nm period were fabricated by a process based on nanoimprint lithographys. Poly (methyl methacrylate) (PMMA) resist was spin-cast on a glass substrate, then imprinted using a SiO\textsubscript{2} mold (50 bar and 170 $^\circ$C for 7 min) and demolded after cooling it down to room temperature. Chromium (Cr) was selectively deposited on each sidewall of the imprinted nano-grating structure by angled deposition. The Cr, deposited on the patterns, induced development of an undercut structure during O\textsubscript{2} reactive ion etching (RIE). This process facilitated the lift-off process and controlled the line-width of the resultant metal grating. O\textsubscript{2} RIE was performed using 10 sccm of O\textsubscript{2} at chamber pressure of 40 mTorr and bias power of 40 W. Ti (2 nm) and gold (50 nm) were deposited using an E-beam evaporator, and the lift-off process provided gold nano-grating patterns. A 2 nm-thick Ti seed layer was used to improve the adhesion of the following gold layer.

\subsection{Stokes polarimetry measurement}
A polarimetric technique was used to measure the shifts of the beam transmitted through the vHMM. We employed two diode lasers, one with a wavelength $\lambda = 638$ nm (Thorlabs, L638P040) and one with $\lambda = 532$ nm (Thorlabs, CPS532). The beam was linearly polarized after a Glan-Tompson polarizer (Thorlabs, GL15) and illuminated to the vHMM at various tilted angles. The beam was captured by a CCD camera (Thorlabs, DCC1545M) after passing through a quarter-wave plate (Thorlabs, AQWP10M-580) and another Glan-Tompson polarizer.







\bibliography{achemso-demo}

\end{document}